# Landau Quantization and Quasiparticle Interference in the Three-Dimensional Dirac Semimetal $Cd_3As_2$

Sangjun Jeon[1*], Brian B. Zhou[1*], Andras Gyenis[1], Benjamin E. Feldman[1], Itamar Kimchi[2], Andrew C. Potter[2], Quinn D. Gibson[3], Robert J. Cava[3], Ashvin Vishwanath[2] and Ali Yazdani[1]

[1]Joseph Henry Laboratories & Department of Physics, Princeton University, Princeton, New Jersey 08544, USA

[2]Department of Physics, University of California, Berkeley, California 09460, USA

[3]Department of Chemistry, Princeton University, Princeton, New Jersey 08544, USA

* These authors contributed equally to this work.

Condensed matter systems provide a rich setting to realize Dirac[1] and Majorana[2] fermionic excitations, as well as the possibility to manipulate them for potential applications[3,4]. It has been recently proposed that chiral, massless particles known as Weyl fermions can emerge in certain bulk materials[5,6] or in topological insulator multilayers[7], and give rise to unusual transport properties, such as charge pumping driven by a chiral anomaly[8]. A pair of Weyl fermions protected by crystalline symmetry, effectively forming a massless Dirac fermion, has been predicted to appear as low energy excitations in a number of materials termed three-dimensional (3D) Dirac semimetals[9–11]. Here we report scanning tunneling microscopy (STM) measurements at sub-Kelvin temperatures and high magnetic fields on the II-V semiconductor $Cd_3As_2$. We probe this system down to atomic length scales, and show that defects mostly influence the valence band, consistent with the observation of ultra-high mobility carriers in the conduction **band. By combining Landau level spectroscopy and quasiparticle interference (QPI), we distinguish a large spin-splitting of the conduction band in a magnetic field and its extended Dirac-like dispersion above the expected regime. A model band structure consistent with our experimental findings suggests that for a magnetic field applied along the axis of the Dirac points, Weyl fermions are the low-energy excitations in $Cd_3As_2$.**

One starting point for accessing the novel phenomena of Weyl fermions is the identification of bulk materials with 3D Dirac points near which the electronic dispersion is linear in all three dimensions[12] in analogy to 2D Dirac points observed in graphene[1] or topological insulators[13]. With time reversal and inversion symmetries preserved, 3D Dirac points can be formed at the crossing of two doubly degenerate bands and constitute two overlapping Weyl points. However, 3D Dirac points are generally not robust to gapping unless they occur along special high symmetry directions in the Brillouin zone, where the band crossing is protected by crystalline point group symmetry[9–11]. In these 3D Dirac semimetals, individual Weyl nodes can be isolated only by breaking either time reversal or inversion symmetry. Since Weyl nodes are topological objects of definite helicity, acting as either a source or sink of the Berry curvature, they are robust against external perturbation and are predicted to harbor exotic effects, such as Fermi arc surface states[5] and chiral, anomalous magneto-transport[8,14]. These unusual transport phenomena of Weyl fermions have been proposed as the basis for novel electronic applications[15].

Several candidate materials, including $Na_3Bi$ and $Cd_3As_2$, were recently predicted[10,11] to exhibit a bulk 3D Dirac semimetal phase with two Dirac points along the $k_z$ axis, stabilized by discrete rotational symmetry. While photoemission measurements[16–19] indeed observed conical dispersions away from certain points in the Brillouin zone of these materials, high energy resolution, atomically-resolved spectroscopic measurements are needed to isolate the physics near the Dirac point and clarify the effect of material inhomogeneity on the low-energy Dirac behavior. Low-temperature scanning tunneling microscopy experiments are therefore ideally suited to address these crucial details. Previously, $Cd_3As_2$ has drawn attention for device applications due to its extremely high room temperature electron mobility[20] (15,000 $cm^2$/Vs), small optical band gap[20], and magnetoresistive properties[21]. The recent recognition that inverted band ordering driven by spin-orbit coupling can foster nontrivial band topology renewed interest in $Cd_3As_2$, which is the only $II_3$-$V_2$ semiconductor believed to have inverted bands. Updated *ab initio* calculations predict 3D Dirac points formed by band inversion between the conduction *s*-states, of mainly Cd-5*s* character, and the heavy hole *p*-states, of mainly As-4*p* character[11,22]. However, the large unit cell of $Cd_3As_2$ with up to 160 atoms due to Cd ordering in a distorted anti-fluorite structure[22] present complications to first-principles calculations, which must be corroborated by careful experimental measurement of the band structure.

To probe the unique electronic structure of $Cd_3As_2$, we perform measurements in a home-built low-temperature STM[23] capable of operating in magnetic fields up to 14 T. Single crystal $Cd_3As_2$ samples are cleaved in ultra-high vacuum and cooled to an electron temperature of 400 mK, where all spectroscopic measurements described here are performed. Figure 1a and its inset show an atomically-ordered topography of a cleaved surface and its associated discrete Fourier transform (DFT). The pseudo-hexagonal Bragg peaks, circled in red, reveal a nearest-neighbor atomic spacing of 4.4±0.15 Å. Their magnitude and orientation precisely match the As-As or Cd-Cd spacing in the (112) plane of this structure[22], schematically illustrated in Fig. 1b, and identify this facet as a natural cleavage plane for $Cd_3As_2$. Because we image atoms at ~96% of the sites in the pseudo-hexagonal lattice, we further attribute the cleaved surface to an As layer, since any Cd layer would contain on average 25% empty sites in this projection.

We present in Fig. 1c the tunneling differential conductance (proportional to the local DOS) measured at B = 0 T along a line spanning 30 nm (see Supplementary Fig. S1a for the precise topographic registry of the linecut). Photoemission measurements[17,18] locate the Dirac point ($E_{Dirac}$) for similar $Cd_3As_2$ samples at -200±20 mV, corresponding to a carrier concentration $n_e$ ~ 2*10^18 cm$^{-3}$. In agreement, the STM conductance spectra show a depression near this energy, and the measured DOS rises as $(E - E_{Dirac})^2$ away from it as expected for 3D Dirac points[24]. The conductance near the Dirac point is nonzero and smooth, representative of a semi-metallic band crossing rather than a band gap. While the presence of surface states can mask a bulk gap, we rule out this possibility by performing QPI measurements, shown below, that do not resolve a strong surface state signal near $E_{Dirac}$. The absence of a gap, particularly at the low temperature of our measurement, is consistent with the proposed theoretical description shown in the inset of Fig. 1c and Fig. 1d, which illustrate a shallow inversion between the valence and conduction bands. Additionally, the zero-field spectra in Fig. 1c display significant spatial fluctuation for energies below $E_{Dirac}$, while in contrast, they are highly homogeneous for energies above $E_{Dirac}$. Since the carrier concentration in as-grown $Cd_3As_2$ is attributed to As vacancies[25], these lattice defects would be expected to primarily impact the valence band rather than the conduction band. In Supplementary Section S1, we show that a common, clustered defect in the As plane (visible as the dark depressions in Fig. 1a) produces strong fluctuations in the conductance of the valence band, but is virtually invisible at the Fermi level. This microscopic information may explain the broad valence band seen in photoemission measurements[17,18] and the high mobility at the Fermi level[20], and suggests routes for further materials optimization.

Landau level spectroscopy with STM has previously been applied to extract precise band structure information for graphene[26], semiconductor 2D electron gases[27], and topological insulator surface states[28,29]. Here in distinction, we extend this technique to quantify the bulk 3D dispersion of $Cd_3As_2$ by applying a magnetic field perpendicular to the cleaved (112) surface of the sample. The 3D band structure is quantized by the magnetic field into effectively 1D Landau bands that disperse along the momentum $k_3$ parallel to the field. The projected bulk DOS measured by STM integrates over all $k_3$ and accordingly displays peaks at the minimum or maximum energies of these Landau bands, which contribute inverse square root divergences to the DOS. Semi-classically, these extrema describe Landau orbits along the constant energy contours of the band structure with extremal cross-sectional area perpendicular to the magnetic field. In Fig. 1e, we illustrate the extremal contours parallel to the (112) plane in $Cd_3As_2$ for energies above and below the Lifshitz transition, demonstrating the merging of two Dirac pockets into a single ellipsoidal contour.

Figure 2a illustrates the Landau level fan diagram for $Cd_3As_2$ assembled from spectra measured from 0 to 14 T at a single fixed location on the sample surface. Four aspects are immediately striking. First, the Landau levels emanate from a point slightly below -200 mV, revealing the presence of a band extremum in the vicinity of the Dirac point determined by photoemission. This suggests that the band inversion is small, consistent with *ab initio* predictions. Second, all prominent Landau levels are electron-like, dispersing towards positive energies with increasing field. The observation of hole-like levels in the valence band is apparently hindered by their electronic disorder, as demonstrated in Fig. 1c, and by their lower band velocity. The data also reveal that the spacing of the Landau levels decreases with Landau level index *n*, indicating a non-parabolic conduction band. Finally, satellite peaks for the dominant Landau level peaks are resolved at high field, revealing the lifting of a degeneracy with increasing field. Figure 2b shows individual spectra for the higher fields which resolve a double peak structure for up to the first 8 pairs of levels (e.g. 12 T).

We first extract information about the band structure of $Cd_3As_2$ from Landau level spectroscopy measurements using a model-independent method. The semi-classical Lifshitz-Onsager relation specifies that the extremal area $S_n$ in reciprocal space for the Landau level *n* occurring at energy $E_n$ must quantize as $S_n = 2\pi e(n + \gamma)B/\hbar$, where $\gamma$ is the phase offset of the quantum oscillations[30]. As verified by QPI measurements presented later, the constant energy contours in the (112) plane are nearly circular; hence, we can take $S_n = \pi k_n^2$, where $k_n$ is the geometric mean of the high symmetric axes of the Fermi surface contour in the (112) plane. We use $\gamma = 1/2$ and adopt an intuitive assignment of the index *n* to the peaks, labeling every two with the same index starting with *n=0* as shown in Fig. 2b (see Supplementary Section II for further discussion). In Fig. 2c, the average peak position $E_n$ and its associated $k_n$ for various B fields trace out an effective dispersion relation. Remarkably, the entire set of peaks collapses onto a single Dirac-like $\sqrt{(n+\gamma)B} \propto |k|$ scaling for a wide energy range, revealing the strong linearity of the conduction band. The linear dispersion with very high Fermi velocity $v_F = 9.4 \pm 0.15 * 10^5$ m/s extends to at least 0.5 V above $E_{Dirac}$, far beyond the expected Lifshitz transition where the two Dirac cones merge. While this extended linearity is not guaranteed by the Dirac physics around the band inversion, it presents important consequences for transport properties of samples with similar carrier concentration. For example, under Boltzmann transport theory for scattering from a screened Coulomb potential, the mobility for a 3D linear dispersion scales as $v_F^2 n_e^{1/3}/n_i$, in stark contrast to the $n_e/(m^{*2} n_i)$ scaling for a 3D quadratic dispersion, where $m^*$ is the effective mass, $n_e$ is the carrier density, and $n_i$ is the concentration of scattering centers (see Ref. 1 for 2D case). This contrasting physical regime for $Cd_3As_2$, which cannot be considered as the limit of normal band structures, may be critical to understanding the ultrahigh mobility and large magnetoresistance reported in a recent transport experiment[31]. Finally, we observe that the extrapolated crossing point from the high energy dispersion occurs at -300 mV, below $E_{Dirac}$, and that the effective velocities of the *n*=0, 1 levels become increasingly small relative to the high energy behavior. We will explain below in detailed modeling that this deviation is a consequence of our sensitivity to the band minimum in the $k_z$ dispersion.

Next, we discuss the spatial homogeneity of the Landau levels. In Fig 2d, we verify that the dominant peak positions are homogeneous in space, with exception of fine features which occur near the Fermi energy. In Fig. 2e and f, we show the *n*=4 and *n*=5 Landau levels for the respective fields when they approach and pass the Fermi level. Remarkably, in certain locations (see Supplementary Fig. S3c), we resolve a four-peak structure in the *n*=5 level and weaker hints of splitting of the *n*=4 state. Because this fine structure occurs in the vicinity of the Fermi level, we speculate that it may arise from band structure effects (states at different momenta but the same energy) that become resolvable near the Fermi level due to the extended electron lifetime, or from many body effects[29]. Since the four-fold structure shifts together with increased field as shown in Fig. 2e and f, we rule out half-filling of the Landau levels[26]. As we are above the Lifshitz transition in this energy range, the additional splitting should also not be interpreted as the lifting of the valley degeneracy of the two Dirac points[26].

Moreover, the spatial resolution of STM enables independent confirmation of the band structure derived from our Landau level spectroscopy measurements. The Fourier transform of spatial modulations in the local DOS mapped by STM provides information about quasiparticle interference (QPI) caused by elastic scattering wavevectors that connect points on the constant energy contour (see Supplementary Section III for discussion of QPI for a 3D band structure). In Fig. 3a-c, we display spectroscopic maps measured at $B = 0\ T$ and $T = 2\ K$ for three different energies that display vivid wave-like features. The evolution of the QPI maps from $E = 450\ mV$ to $E = 150\ mV$ shows the length of the scattering wavevectors to increase with decreasing energy. At $E = -200\ mV$, this interference signal can no longer be resolved as the diverging wavelength near $E_{Dirac}$ overlaps with the background electronic puddling. The interference patterns seen in the DFTs of the spectroscopic maps (see Fig. 3d-f and Supplementary Fig. S5) distinguish the shape of the extremal Fermi contours perpendicular to [112] above the Lifshitz transition as quasi-circular, justifying our previous assumption. Figure 3g demonstrates the consistency between the extracted QPI dispersion and the semi-classical Landau level analysis, which together resolve a conduction band that onsets near -200 mV and disperses linearly to high energies. Above 500 mV, the linear dispersion becomes flatter and a second scattering vector, likely from another bulk band, is resolved in the QPI data.

To gain further insight into the nontrivial Landau level structure and to determine when Weyl fermions appear as the low energy excitations of $Cd_3As_2$, we introduce a band structure model that captures the salient features of our data. Following previous work, the low-energy dispersion around the Γ point for $Cd_3As_2$ can be described by an inverted HgTe-type band model using a minimal 4-band basis of the $\left|S_{\frac{1}{2}}, \frac{1}{2}\right\rangle$, $\left|P_{\frac{3}{2}}, \frac{3}{2}\right\rangle$, $\left|S_{\frac{1}{2}}, -\frac{1}{2}\right\rangle$, and $\left|P_{\frac{3}{2}}, -\frac{3}{2}\right\rangle$ states [11];

$$H_{eff}(\vec{k}) = \varepsilon_o(\vec{k}) + \begin{pmatrix} M(\vec{k}) & Ak_+ & 0 & 0 \\ Ak_- & -M(\vec{k}) & 0 & 0 \\ 0 & 0 & M(\vec{k}) & -Ak_- \\ 0 & 0 & -Ak_+ & -M(\vec{k}) \end{pmatrix},$$

where $\varepsilon_o(\vec{k})$ and $M(\vec{k})$ encode the band structure and $k_\pm = k_x \pm ik_y$ (details are described in Supplementary Section IV). Landau quantization in the (112) plane reflects a mixture of the $k_x$-$k_y$ and $k_z$ dispersions. Hence, the linearity in Fig. 2c implies that both the $k_x$-$k_y$ and $k_z$ dispersion are linear at high energies. To capture this trend, we modify the original parabolic $k_z$ dispersion in $M(\vec{k})$ to be hyperbolic. This simple modification maintains all qualitative aspects of the low energy band inversion and is essential for modeling the extended energy range of the data. When a magnetic field is applied, we transform the momentum $\vec{k} \to \vec{k} - \frac{e}{\hbar}\vec{A}$ via Peierls substitution of the magnetic vector potential $\vec{A}$ and include a Zeeman term in the total Hamiltonian $H(\vec{k}) = H_{eff}(\vec{k}) + H_{Zeeman}(\vec{k})$.

In Fig. 4a, we show the results of numerical Landau level simulations using band structure parameters consistent with the $k_x$-$k_y$ dispersion measured by photoemission[17,18] and with the presence of band inversion indicated by our zero field spectra. Although a precise determination of the size of the inversion is not possible (20 mV is used in Fig. 4), the data are more consistent with shallower band inversions. Nevertheless, the model illustrates the essential physical origin for the observed Landau level structure. At high fields, the DOS singularities observed in the data correspond to the energies of the Landau level band minima at the Γ point (see Supplementary Section V for discussion of the low field regime where additional extrema may occur inside the two Dirac cones). Hence, the deviation from Dirac scaling for the lowest levels in Fig. 2c reflects the parabolic (massive) band minimum in the $k_z$ dispersion, which is probed by the tilted magnetic field.

More importantly, the agreement of our data with the model calculations suggests that the Landau level doublet structure arises from a combination of orbital and Zeeman splitting of the spin-degenerate conduction band. Orbital splitting depends on the shape of the band structure and diminishes away from $E_{Dirac}$. In Fig. 4b, we theoretically illustrate this evolution of the Landau levels due to orbital effects as the angle of the field is tilted away from the c axis (for clarity we have set the Zeeman term to zero here as it introduces only an additional nearly constant splitting). For our data, measured at the intermediate angle denoted by the yellow bar, it is natural to adopt the assignment scheme *n* shown on the right side of Fig. 4b such that the pairs of levels closest in energy have the same index. In Fig. 4c, we extract an effective total $g^*$ from the experimental Landau level splitting for each index at several different magnetic fields. We find that $g^*=37\pm2$ for the lowest level and that $g^*$ decreases with increasing energy from $E_{Dirac}$, consistent with theoretical models based on prior Shubnikov-de-Haas measurements[32].

In the case of a magnetic field titled from the *c* axis, calculations based on our model band structure show that the Weyl nodes are eliminated by small gaps at the Dirac points caused by the broken rotational ($C_4$) symmetry (Fig. 4d). Therefore, to observe Weyl fermions, application of a magnetic field along [001] is required to break time reversal symmetry while maintaining $C_4$ symmetry (Fig. 4e). Moreover, the direction of the magnetic field is shown here to tune the orbital and orbital-independent splitting in in $Cd_3As_2$. Exploration of that phase space in samples with lower carrier concentration opens the possibility of engineering and observing topological states in 3D Dirac materials.

*Note added:* After submission of our manuscript, we became aware of Ref. 33 which reported that the two Dirac points for the (112) cleave of $Cd_3As_2$ occur along the [112] direction. The shift of the Dirac points from the [001] direction as in previous bulk calculations (Ref. 11) was attributed to the assumed absence of Cd ordering in the surface layers. Differences in the measurement results of their samples and ours (see also Ref. 18 and 31), such as the carrier concentration, band velocities, and location of the Dirac points, may stem from differences in the details of the crystal structure.

**Methods**

$Cd_3As_2$ crystals were grown from a Cd rich melt with the stoichiometry $Cd_8As_3$ in an evacuated quartz ampoule. The sample was heated to 800 °C at 3 °C/min, then slowly cooled to 400 °C at 0.1 °C/min. After reaching 400 °C, the sample was subsequently centrifuged to remove excess molten Cd. Multiple samples (1 mm x 1 mm x 0.5 mm) were cleaved at room temperature parallel to the largest flat face in ultra-high vacuum conditions, producing the (112) cleavage plane each time, and subsequently cooled down to the electron temperature of 400 mK at which point spectrum measurements were performed with a platinum-iridium tip. QPI maps were measured at 2 K with the same STM setup. dI/dV spectra were acquired using lock-in amplifier at a frequency of 798.7 Hz.

**Acknowledgement**

The work at Princeton and the Princeton Nanoscale Microscopy Laboratory was supported by the ARO MURI program W911NF-12-1-0461, DARPA-SPWAR Meso program N6601-11-1-4110, NSF-DMR1104612, ONR- N00014-11-1-0635, and NSF-MRSEC NSF-DMR0819860 programs.

**Author contributions**

S.J. and B.B.Z performed STM experiments with assistance from A.G. Theoretical simulations were constructed by I.K., A.C.P., and A.V. Q.D.G. and R.J.C. synthesized the materials. S.J. and B.B.Z. performed analysis and modeling. The manuscript was written by S.J., B.B.Z., B.E.F., and A.Y. All authors commented on the manuscript. Correspondence and requests for materials should be addressed to A.Y. (yazdani@princeton.edu)

**References**

1. Castro Neto, a. H., Peres, N. M. R., Novoselov, K. S. & Geim, a. K. The electronic properties of graphene. *Rev. Mod. Phys.* **81,** 109–162 (2009).

2. Beenakker, C. W. J. Search for Majorana Fermions in Superconductors. *Annu. Rev. Condens. Matter Phys.* **4,** 113–136 (2013).

3. Novoselov, K. S. *et al.* A roadmap for graphene. *Nature* **490,** 192–200 (2012).

4. Nayak, C., Stern, A., Freedman, M. & Das Sarma, S. Non-Abelian anyons and topological quantum computation. *Rev. Mod. Phys.* **80,** 1083–1159 (2008).

5. Wan, X., Turner, A. M., Vishwanath, A. & Savrasov, S. Y. Topological semimetal and Fermi-arc surface states in the electronic structure of pyrochlore iridates. *Phys. Rev. B* **83,** 205101 (2011).

6. Xu, G., Weng, H., Wang, Z., Dai, X. & Fang, Z. Chern Semimetal and the Quantized Anomalous Hall Effect in $HgCr_2Se_4$. *Phys. Rev. Lett.* **107,** 186806 (2011).

7. Burkov, A. A. & Balents, L. Weyl Semimetal in a Topological Insulator Multilayer. *Phys. Rev. Lett.* **107,** 127205 (2011).

8. Hosur, P. & Qi, X. Recent developments in transport phenomena in Weyl semimetals. *Comptes Rendus Phys.* **14,** 857–870 (2013).

9. Young, S. M. *et al.* Dirac Semimetal in Three Dimensions. *Phys. Rev. Lett.* **108,** 140405 (2012).

10. Wang, Z. *et al.* Dirac semimetal and topological phase transitions in $A_3Bi$ (A=Na, K, Rb). *Phys. Rev. B* **85,** 195320 (2012).

11. Wang, Z., Weng, H., Wu, Q., Dai, X. & Fang, Z. Three-dimensional Dirac semimetal and quantum transport in $Cd_3As_2$. *Phys. Rev. B* **88,** 125427 (2013).

12. Orlita, M. *et al.* Observation of three-dimensional massless Kane fermions in a zinc-blende crystal. *Nat. Phys.* **10,** 233 (2014).

13. Hasan, M. Z. & Kane, C. L. Colloquium: Topological insulators. *Rev. Mod. Phys.* **82,** 3045–3067 (2010).

14. Potter, A. C., Kimchi, I. & Vishwanath, A. Quantum Oscillations from Surface Fermi-Arcs in Weyl and Dirac Semi-Metals. *arXiv*:1402.6342 (2014).

15. Kharzeev, D. E. & Yee, H.-U. Anomaly induced chiral magnetic current in a Weyl semimetal: Chiral electronics. *Phys. Rev. B* **88,** 115119 (2013).

16. Liu, Z. K. *et al.* Discovery of a Three-dimensional Topological Dirac Semimetal , $Na_3Bi$. *Science* **343,** 864–867 (2014).

17. Neupane, M. *et al.* Observation of a three-dimensional topological Dirac semimetal phase in high-mobility $Cd_3As_2$. *Nat. Commun.* **5,** 3786 (2014).

18. Borisenko, S., Gibson, Q., Evtushinsky, D., Zabolotnyy, V. & Büchner, B. Experimental Realization of a Three-Dimensional Dirac Semimetal. *arXiv*:1309.7978 (2013).

Page **14** of **17**

19. Xu, S.-Y. *et al.* Observation of a bulk 3D Dirac multiplet, Lifshitz transition, and nestled spin states in Na$_3$Bi. *arXiv* :1312.7624 (2013).

20. Turner, W., Fischler, A. & Reese, W. Physical properties of several II-V semiconductors. *Phys. Rev.* **121,** 759 (1961).

21. Radautsan, S. I., Arushanov, E. K. & G. P. Chuiko. The conduction band of cadmium arsenide. *Phys. Status Solidi* **20,** 221–226 (1973).

22. Ali, M. N. *et al.* The crystal and electronic structures of Cd3As2, the 3D electronic analogue to graphene. *Inorg. Chem.* **53,** 4062 (2014).

23. Misra, S. *et al.* Design and performance of an ultra-high vacuum scanning tunneling microscope operating at dilution refrigerator temperatures and high magnetic fields. *Rev. Sci. Instrum.* **84,** 103903 (2013).

24. Ashby, P. E. C. & Carbotte, J. P. Theory of magnetic oscillations in Weyl semimetals. *Eur. Phys. J. B* **87,** 92 (2014).

25. Spitzer, D. P., Castellion, G. A. & Haacke, G. Anomalous Thermal Conductivity of Cd$_3$As$_2$ and the Cd$_3$As$_2$−Zn$_3$As$_2$ Alloys. *J. Appl. Phys.* **37,** 3795 (1966).

26. Song, Y. J. *et al.* High-resolution tunnelling spectroscopy of a graphene quartet. *Nature* **467,** 185–9 (2010).

27. Morgenstern, M., Klijn, J., Meyer, C. & Wiesendanger, R. Real-Space Observation of Drift States in a Two-Dimensional Electron System at High Magnetic Fields. *Phys. Rev. Lett.* **90,** 056804 (2003).

28. Okada, Y., Serbyn, M., Lin, H. & Walkup, D. Observation of Dirac node formation and mass acquisition in a topological crystalline insulator. *Science* **341,** 1496–1499 (2013).

29. Hanaguri, T., Igarashi, K. & Kawamura, M. Momentum-resolved Landau-level spectroscopy of Dirac surface state in Bi$_2$Se$_3$. *Phys. Rev. B* **82,** 081305 (2010).

30. Wright, A. R. & McKenzie, R. H. Quantum oscillations and Berry's phase in topological insulator surface states with broken particle-hole symmetry. *Phys. Rev. B* **87,** 085411 (2013).

31. Liang, T. *et al.* Ultrahigh mobility and giant magnetoresistance in Cd$_3$As$_2$ : protection from backscattering in a Dirac semimetal. *arXiv* :1404.7794 (2014).

32. Wallace, P. R. Electronic g-Factor in Cd$_3$As$_2$. *Phys. Stat. Sol.* **92,** 49–55 (1979).

33. Liu, Z. K. *et al.* A stable three-dimensional topological Dirac semimetal Cd3As2. *Nat. Mater.* Advanced online publication, May 25 (2014). doi:10.1038/NMAT3990

**Figure captions:**

**Fig.1: Crystal and band structures of Cd$_3$As$_2$ (112) cleaved crystal. a,** Atomically-ordered topographic image (I=50pA, V=-250mV) of the Cd$_3$As$_2$ (112) surface. Inset shows its 2D Fourier transform. Red circles are associated with Bragg peaks and blue circles with reconstruction peaks. **b,** Schematic of the Cd$_3$As$_2$ unit cell along the (112) plane (red). Cd atoms and As atoms both make a pseudo-hexagonal lattice. **c,** Differential conductance spectra (I=300pA, V=250mV) taken at 90 spatial positions over a line spanning 30 nm. The blue curves show the individual spectra and the red curve is the spatial average. Spatial variation in the local density of states is especially pronounced below the Dirac point. The inset shows the schematic band dispersion along the [001] direction passing the Γ point. **d,** Schematic band structure of Cd$_3$As$_2$ based on *ab initio* calculations. Two 3D Dirac points marked as $k_D^+$, $k_D^-$ are located along the [001] direction and are evenly separated from the Γ point. The $k_\perp$ direction refers to any axis perpendicular to the $k_z$ direction. **e,** Schematic of the Fermi surfaces above (red) and below (blue) the Lifshitz transition. The overlaid solid curves represent the extremal cross-sections parallel to the (112) plane, showing two pockets merging into a single ellipsoidal contour. Throughout this paper, $k_x$, $k_y$, $k_z$ are parallel to the *a, b, c* axes, respectively, of the unit cell denoted in **b**.

**Fig.2: Landau level spectroscopy. a,** Landau level fan diagram measured at 400 mK (I=400pA, V=-250mV, $V_{osc}$=0.8mV), consisting of point spectra in 1T increments. The variation of the spatial position for each spectrum is less than 0.5 nm. **b,** Splitting of Landau Levels. The point spectra (I=500pA, V=300mV, $V_{osc}$=1.2mV), obtained from 10T to 14T in 0.5 T increments, show a doublet Landau peak structure whose separation decreases at high index. Plots are shifted vertically and a smooth background is subtracted based on the 2T data. **c,** Effective band dispersion in the (112) plane formulated from the Lifshitz-Onsager quantization condition. Sixteen Landau levels for each magnetic field are plotted, where the average energy is used for indexes with two split peaks. The black line is the linear extrapolation of the velocity at the Fermi level. **d,** Spatial variation of Landau Levels at 12.25 T (I=400pA, V=-250mV, $V_{osc}$=1 mV) ranging from -250 mV to 100 mV. The green curve is the spatial average. Landau levels are homogeneous in space except around the Fermi level. **e,f,** Spatial variation of Landau Levels around the Fermi level at 12.25 T and 12.75 T (I=400pA, V=-250mV, $V_{osc}$=0.3 mV). The spectra in **d, e,** and **f** were all taken along the same line cut. Fine features, such as the 4-peak structure of the n=5 level observed at certain locations and weakly in the spatial average, are visible only around the Fermi level.

**Fig.3: Bulk quasi-particle interference projected onto (112) plane. a-c,** Spectroscopic maps of $Cd_3As_2$ at 450 meV, 150 meV, and -200 meV, respectively. Strong interference features are visible at 450 meV, while only electronic puddles are observed at -200 meV. **d-f,** 2D discrete Fourier transforms (2D-DFTs) of **a, b,** and **c,** respectively. The red dashed circles show the scattering of the electron-like conduction

band ($\Lambda_6$ band), and the cyan dashed circle shows that of a second band which emerges at higher energy. **g,** Plot of QPI peaks and reproduced Landau level peaks. The red (cyan) momentum vectors are obtained from the radius of the QPI feature labeled by the red (cyan) dashed circle in **d** and **e**. Blue and green curves are guides to the eye. The orange circles reproduce the Landau level data shown in Fig. 2c.

**Fig.4: Landau level simulation with modified Kane Hamiltonian. a,** Simulation of Landau levels and their splitting (peak positions of 10T to 14T Landau level spectra are plotted as red circles). The electron-like (blue curves) and hole-like levels (red curves) are derived from the extrema of the Landau level bands at the Γ point. The Dirac and Lifshitz points at zero field are marked on the vertical axis. **b,** Theoretical angle-dependent orbital splitting of the Landau levels. The measurements reported here were performed at θ=54.7°, denoted by the yellow bar. **c,** Effective total g-factor g* extracted from the experimental data as a function of energy and magnetic field. **d,e,** Calculated Landau level bands for a magnetic field along the [112] direction and [001] direction. The corresponding density of states is shown for the [112] directed field. The inset in (**d**) and (**e**) zooms in on the crossing point between the lowest electron- and hole-like bands, showing the opening of a gap in (**d**) due to broken $C_4$ symmetry.

**Fig. 1**

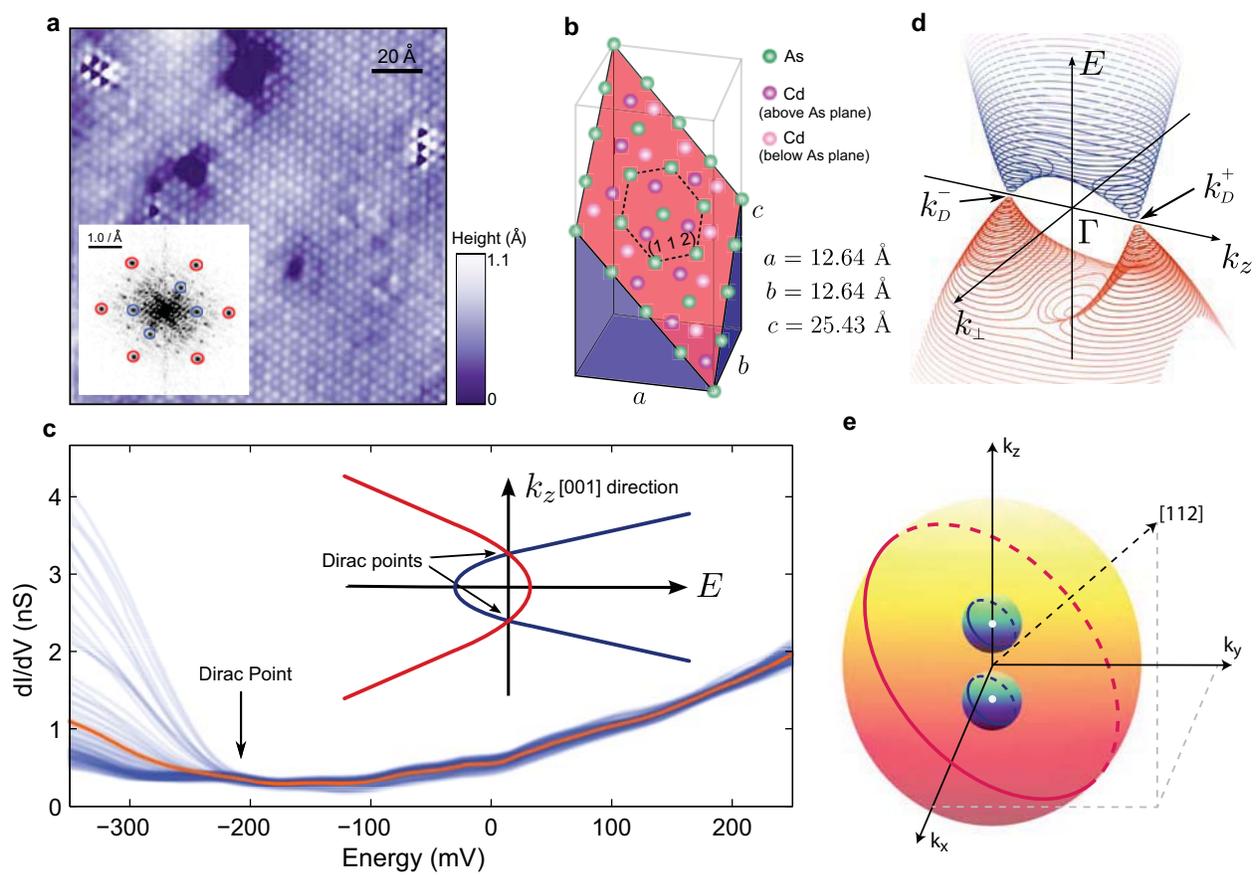

**Fig. 2**

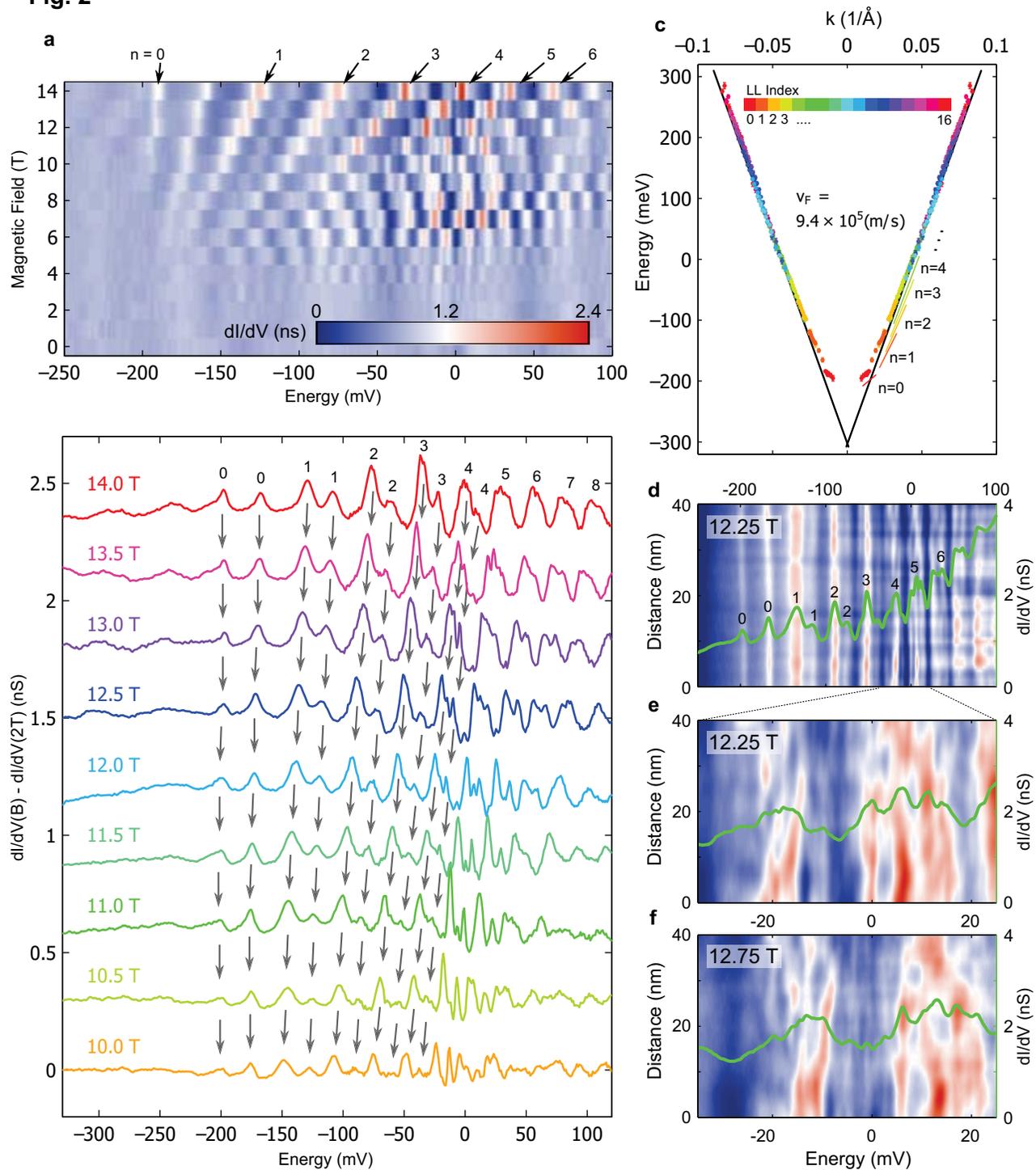

**Fig. 3**

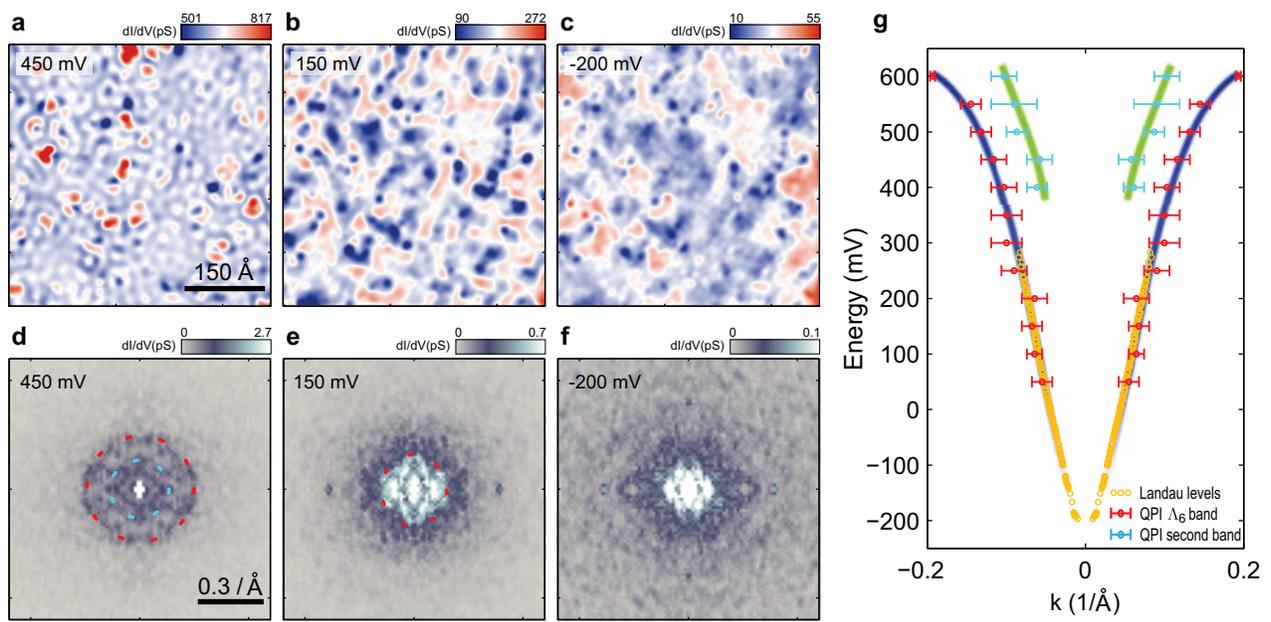

**Fig. 4**

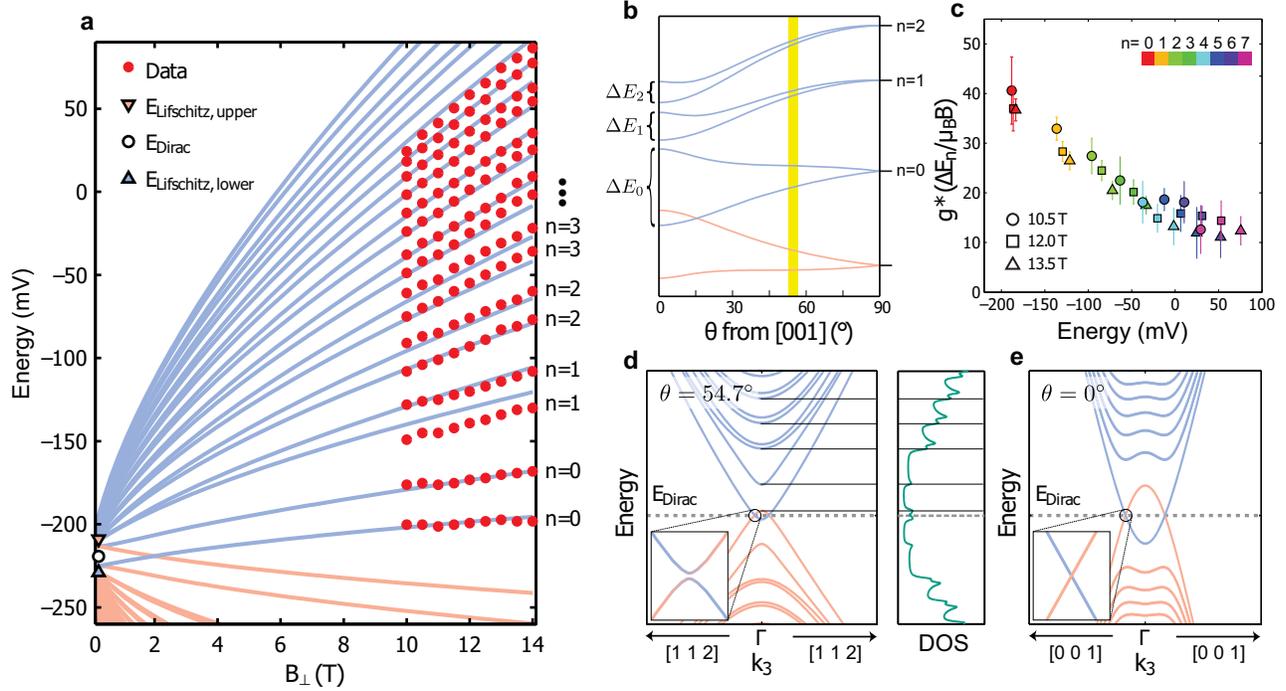

# Landau Quantization and Quasiparticle Interference in the Three-Dimensional Dirac Semimetal $Cd_3As_2$

Sangjun Jeon[1*], Brian B. Zhou[1*], Andras Gyenis[1], Benjamin E. Feldman[1], Itamar Kimchi[2], Andrew C. Potter[2], Quinn D. Gibson[3], Robert J. Cava[3], Ashvin Vishwanath[2] and Ali Yazdani[1]

[1]Joseph Henry Laboratories & Department of Physics, Princeton University, Princeton, New Jersey 08544, USA
[2]Department of Physics, University of California, Berkeley, California 09460, USA
[3]Department of Chemistry, Princeton University, Princeton, New Jersey 08544, USA
* These authors contributed equally to this work.

## Section I. Spatial variation of the conductance spectra

We present the spatial variation of the conductance spectra with and without applying a magnetic field. Figure S1a shows the topography data where the local conductance spectra are measured consecutively along the red line at 0T (Fig. 1c) and 14T (Fig. S1b). As discussed in Fig. 2d in the main text, Landau levels for the electron-like bands are robust over space, while Landau levels for the hole-like bands are absent. The in-field spectra remain spatially inhomogeneous below the Dirac point.

Figure S2a shows a large 500x500 Å$^2$ area of the (112) cleaved surface of $Cd_3As_2$. For the crystal structure of $Cd_3As_2$, both As and Cd planes exist parallel to the (112) plane, but a quarter of the lattice sites in the Cd layers would be empty due to the non-commensurability of the $Cd_3As_2$ formula unit to the ideal anti-fluorite crystal structure. In contrast, the As layer would have a nearly perfect hexagonal lattice structure. While the imaged lattice spacing is consistent with either the As-As or Cd-Cd distance, the topography data are more consistent with the As layer because the number of empty sites is less than 5% of the total number of atoms as shown in Fig. 1a and Fig. S2a. As discussed in the main text, the electron-like band is derived primarily from Cd-*5s* states and hole-like band from As-*4p* states[1]. Moreover, since the carriers in naturally grown $Cd_3As_2$ are primarily donated by As vacancies[2], this type of lattice defect impacts the hole-like band more than the electron-like band. As exemplified by Fig. S2, stronger conductance variation is seen at energies in the hole-like band (e.g. *E = -500 mV* in Fig. S2b), than at energies in the electron-like band (e.g. *E = 0 mV* in Fig. S2c). Surface defects in the As plane imaged as depressions in the topography localize strong enhancements in conductance in the hole-like band, while they negligibly impact the conductance at the Fermi level, which may explain the high electron mobility of $Cd_3As_2$ measured at the Fermi level.

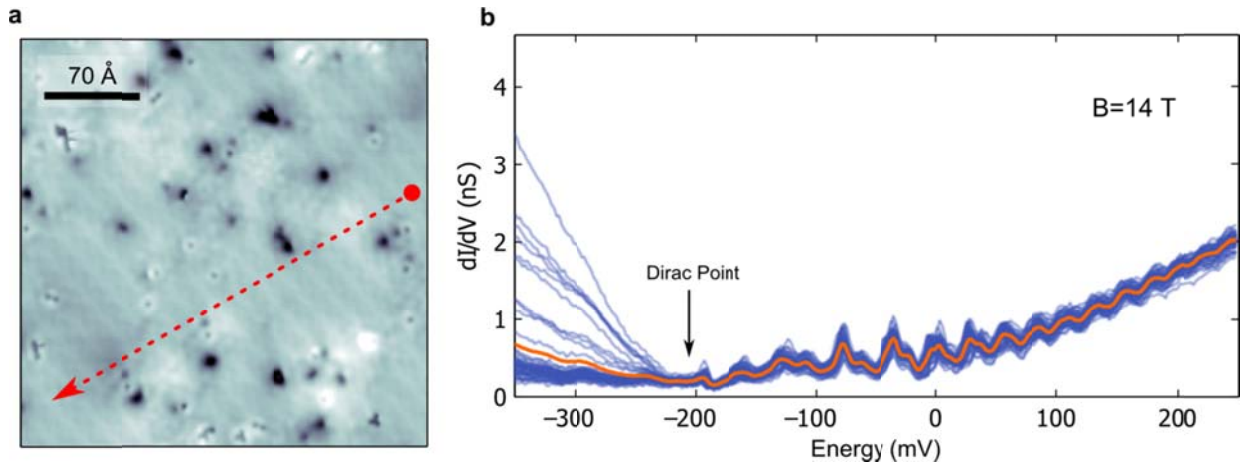

**Figure S1: Spatial variation of Landau Levels at 14 T. a,** The topography data (V=300mV, I=30pA) show surface defects and a reconstruction. **b,** Fifty individual local conductance spectra (I=500pA, V=300mV, $V_{osc}$=1.5mV) are acquired at 14 T along the red dashed line shown in **a**. The blue curves are raw data and the red curve is the spatial average. The spectra in Fig. 1c of the main text are also measured along the same line as in panel **a**.

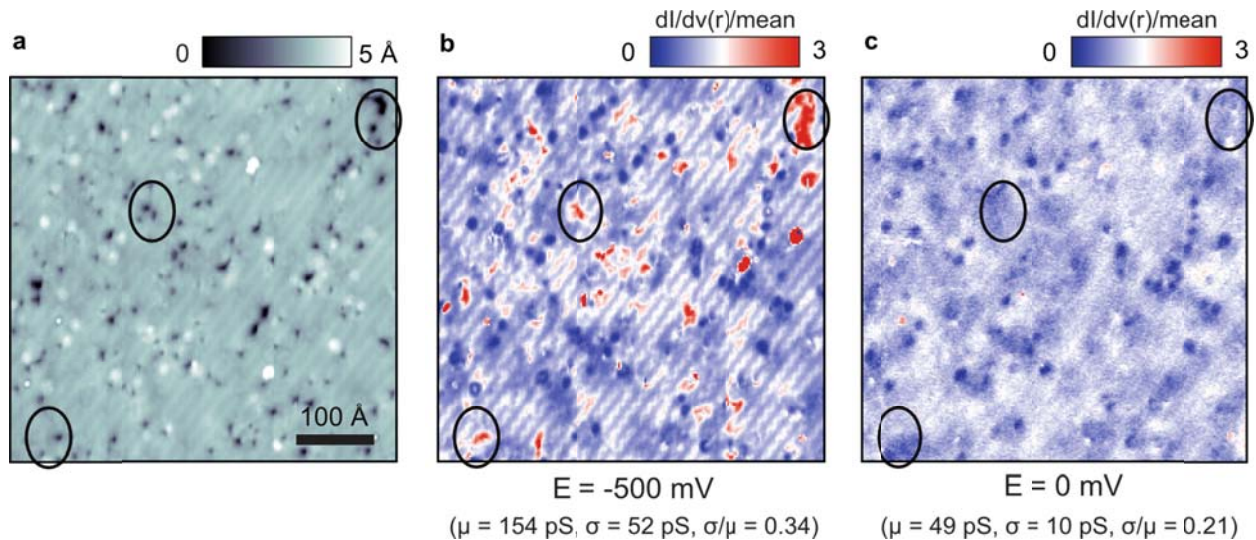

**Figure S2: Spatial variation of the local conductance for the hole- and electron- like bands. a,** The simultaneous topography data (V=600mV, I=300pA) show surface defects. **b,** Conductance map normalized by its mean value measured at -500 mV. The mean value is 154 pS and the standard deviation is 52 pS. **c,** Normalized conductance map measured at the Fermi level. The mean value is 49 pS and the standard deviation is 10 pS. The conductance for the hole-like band (**b**) is enhanced around defects which appear as dark depressions in the topography, but the conductance is minimally changed for the electron band over the same defect (**c**). For comparison, the black circles identify the same locations for each figure. The maps shown in this figure were taken at $B = 14\ T$, but the effect discussed is independent of the field.

Figure S3a shows topography data where the high resolution conductance spectra for Fig. 2d-f are acquired (I=400 pA, V=-250 mV, $V_{osc}$= 0.3 mV). The local density of states around the Fermi level varies with space as shown in Fig. S3b, which reproduces the data in Fig. 2e of the main text. The point spectrum shown for the location denoted by the green circle in Fig. S3a and corresponding to the green dashed line in Fig. S3b is displayed in Fig. S3c. This location displays most prominently the fine structure of the Landau level peaks near the Fermi energy.

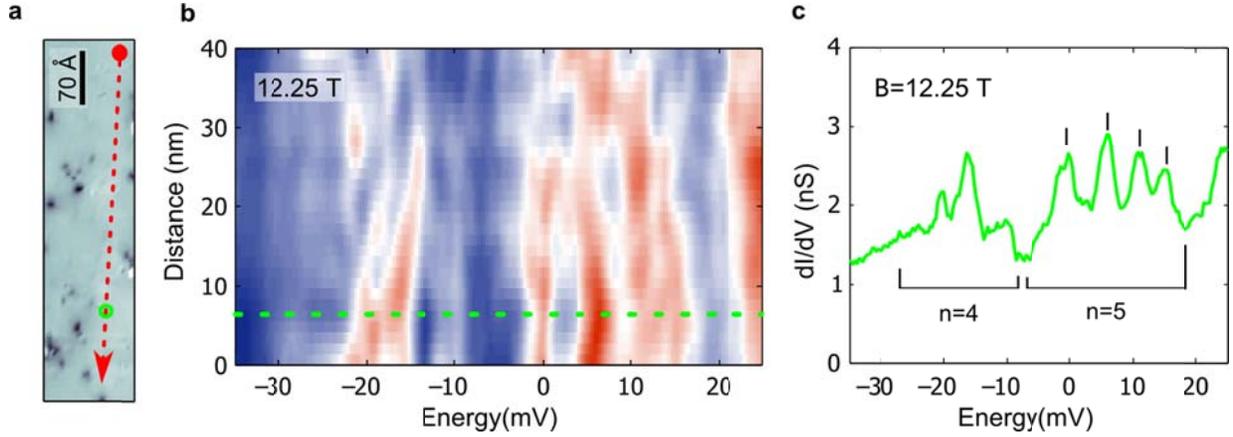

**Figure S3: Spatial variation of fine splitting around the Fermi level at 12.25 T. a**, The topography data shows the location where local conductance spectra are obtained for this figure and also for Fig. 2d-f of the main text. **b**, Local conductance variation along the red line shown in **a**. **c**, Single conductance spectrum at the green circle shown in **a** and corresponding to the green line shown in **b**. The spectrum shows four peaks for the n=5 Landau level.

## Section II. Determining Landau level index and Berry phase

To resolve the ambiguity in determining the orbital index *n* to the electron-like Landau levels, we attempt two possible indexing schemes for the peaks as labeled in the inset of Fig. S4a (starting with *n = 0)* and Fig. S4b (starting with *n = 1*). We plot the effective band dispersion following the same semi-classical procedure described in the main text. The phase factor $\gamma$ is related to the Berry phase $\Phi_B$ enclosed by the cyclotron orbit through $\gamma = \frac{1}{2} - \frac{\Phi_B}{2\pi}$. We use $\gamma = 1/2$ for the plots since the cyclotron orbits at energies above the Lifshitz transition enclose two Dirac points of equal and opposite Berry phase (although at energies near the Lifshitz transition, $\gamma$ may deviate slightly from $1/2$)[3]. With the proper orbital index assignment, Landau level positions for all indexes and magnetic fields should collapse onto a single smooth line because the cross-sectional area of the Fermi surface is a continuous and single-valued function of the energy. Figure S4a shows that all the data points collapse onto a single line within the error bar for the assignment scheme starting with $n = 0$. Thus, we identify this assignment scheme as the correct one for analyzing the Landau level spectra.

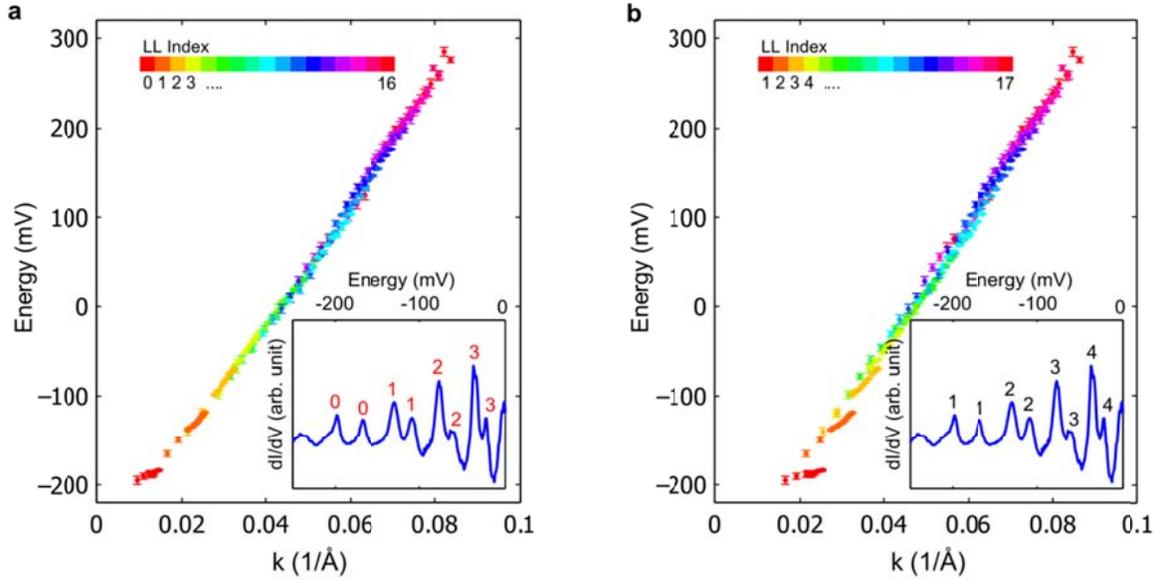

**Figure S4: Landau level index assignment. a,b,** The effective band dispersion formulated from the Lifshitz-Onsager relation with different index assignments described in the inset of **a** and **b**. The spectra shown in the inset is the Landau level spectrum measured at 14T.

## Section III. Symmetrization of the discrete Fourier transform (DFT) of quasiparticle interference (QPI) maps

We perform conductance maps at various energies spanning -200 mV to 600 mV at 0T. The distance between tip and sample is stabilized by setting the tunneling current to 300 pA for a bias of 600 mV at each spatial point in a 208x208 grid over a 500x500 Å² area. The real space conductance map at a specified energy shows wave-like features (Fig. S5a), whose momentum information is obtained from the real space map's two dimensional DFT (Fig. S5b). The QPI for a 3D band structure can be approximated as the integration of the 2D QPI intensities for the Fermi surface planes at fixed $k_3$ perpendicular to the sample surface[4]. Therefore, QPI for the nearly spherical constant energy surfaces of $Cd_3As_2$ should represent the weighted sum of 2D QPI patterns for contours of "latitude" at fixed $k_3$. Each ring of latitude of radius $k$ contributes a QPI ring of radius $2k$; thus, the sum will fill up a disc-like pattern. The regions near the equator of the Fermi surface (maximal $k_{max}$) will be weighted more due to the slower dispersion of the radius of the ring of latitude, while regions near the pole ($k \to 0$) will be weighted less due to the faster dispersion. Consistently, the experimental QPI data contains intensity up to the maximal radius $2k_{max}$, as is evident for E = 300 mV shown in Fig. 5.

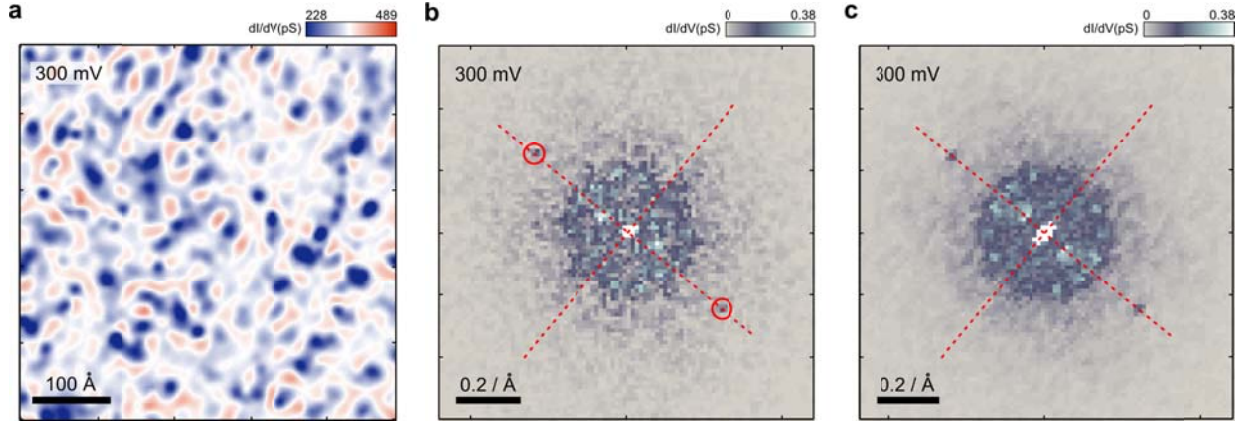

**Figure S5: Symmetrization method for the discrete Fourier transformed (DFT) maps. a,** A conductance map at 300 mV over the topographic area shown in Fig. S2a. **b,** Two dimensional DFT of the conductance map in (**a**) without symmetrization. The peaks marked by the red circles originate from the surface reconstruction. **c,** Two dimensional DFT of the conductance map in (**a**) with symmetrization. Two mirror axes (red lines) for symmetrization are illustrated in panels **b** and **c**.

To enhance the signal-to-noise ratio, we symmetrize the obtained DFT map based on the symmetry of the $Cd_3As_2$ band structure. The atomic resolution topography shown in the Fig.1a reveals that the (112) plane is the cleaved surface. In the (112) plane, there are two symmetric axes, which form mirror planes: one mirror plane is perpendicular to $[\bar{1}\,\bar{1}\,1]$ and another is perpendicular to $[1\,\bar{1}\,0]$. From comparison to the atomic lattice imaged for this sample, we determine that the one-dimensional reconstruction feature (visible in Fig. S2a, which shows the topographic image of the conductance map area, and identified as the features circled in red in the DFT of Fig. S5b) is either along the $[\bar{1}\,\bar{1}\,1]$ or along the $[1\,\bar{1}\,0]$ direction. Therefore we take the axis passing the two reconstruction peaks and its perpendicular axis as the symmetric axes of the DFT map (red lines in the Fig. S5b and c). Figure S5c is the result of symmetrizing the data in Fig. S5b along the two mirror planes. In Fig. 3d-f in the main text, we further rotate the symmetrized map such that the symmetric axis are horizontal and vertical to the axes of the image. The radius $Q$ of the quasi-circular pattern in the DFT map is then converted to the effective momentum $k$ for Fig. 3g by the scattering relation $Q = 2\,k$.

## Section IV. Modified four-band Kane Hamiltonian

In this section, we describe the four-band model Hamiltonian used in this study to capture the Landau level spectra of $Cd_3As_2$ within the framework of prior *ab initio* calculations and ARPES[5,6] measurements. An eight band Kane model is generally used to describe the low-energy band structure of small band gap semiconductors such as InSb and HgTe[7]. Near the charge neutrality point, the description of the 3D Dirac semimetal phase can be further reduced to a Hamiltonian using the four-band basis of the $\left|S_{\frac{1}{2}}, \frac{1}{2}\right\rangle$, $\left|P_{\frac{3}{2}}, \frac{3}{2}\right\rangle$, $\left|S_{\frac{1}{2}}, -\frac{1}{2}\right\rangle$, and $\left|P_{\frac{3}{2}}, -\frac{3}{2}\right\rangle$ states[1]. The total Hamiltonian is composed of two terms

$$H = H_{eff} + H_{Zeeman},$$

where $H_{eff}$ models the band structure and $H_{Zeeman}$ is the Zeeman energy. The effective Hamiltonian $H_{eff}$ has the form

$$H_{eff}(\vec{k}) = \varepsilon_o(\vec{k}) + \begin{pmatrix} M(\vec{k}) & Ak_+ & Dk_- & B^*(\vec{k}) \\ Ak_- & -M(\vec{k}) & B^*(\vec{k}) & 0 \\ Dk_+ & B(\vec{k}) & M(\vec{k}) & -Ak_- \\ B(\vec{k}) & 0 & -Ak_+ & -M(\vec{k}) \end{pmatrix},$$

where $k_{\pm} = k_x \pm i k_y$, and the terms $A$, $\varepsilon_o(\vec{k})$, $M(\vec{k})$ encode the electronic dispersion. We ignore the terms involving $D$, which describe any possible inversion symmetry breaking in the crystal structure, and $B(\vec{k})$, which contains higher order terms allowed by crystal symmetry, since they are expected to produce only higher order corrections. The direction of the momentum $k_x$, $k_y$, and $k_z$ are associated with the *a*, *b*, and *c* axes of the crystal, respectively. However in the presence of a magnetic field, it is more convenient to describe the Hamiltonian in terms of the magnetic axis ($k_3$) parallel to the field, because the momentums $k_1$ and $k_2$ perpendicular to the magnetic axis are quantized by the field. Accordingly, $k_1$ and $k_2$ can be transformed into the sum of ladder operators.

$$k_1 = \frac{i}{\sqrt{2}\ell_B}(a^\dagger - a), \quad k_2 = \frac{1}{\sqrt{2}\ell_B}(a^\dagger + a),$$

where $\ell_B = \sqrt{\frac{\hbar}{eB}}$ is the magnetic length. The raising (lowering) operators $a^\dagger$ ($a$) for the Landau levels $|n\rangle$ obey the usual relations: $a^\dagger|n\rangle = \sqrt{n+1}\,|n+1\rangle$, $a|n\rangle = \sqrt{n}\,|n-1\rangle$. We use a 3D rotation matrix $U$ to transform the vectors $(k_1, k_2, k_3)$ of the magnetic frame into the $(k_x, k_y, k_z)$ of the crystal frame,

$$\begin{pmatrix} k_x \\ k_y \\ k_z \end{pmatrix} = U \begin{pmatrix} k_1 \\ k_2 \\ k_3 \end{pmatrix}$$

and numerically diagonalize the Hamiltonian in the $|n\rangle$ basis to compute the Landau level energy spectrum.

The semi-classical Landau quantization analysis shown in Fig. 2c demonstrates that electronic dispersion in the tilted (112) plane for the conduction band of $Cd_3As_2$ is linear in an extended energy range from -100 mV to 300 mV. The original Hamiltonian proposed in Ref. S1 allows for both linear (via $Ak_\pm$) and quadratic terms in the dispersion in the $k_x$ or $k_y$ directions, but only a quadratic term in the $k_z$ direction. To reflect the linear dispersion in a wide energy range in all three momentum directions, we introduce a hyperbolic dispersion along the $k_z$ direction.

$$\varepsilon_o(\vec{k}) = C_0 + C_1 k_z^2 + C_2 (k_x^2 + k_y^2)$$

$$M(\vec{k}) = M_0 + \sqrt{M_3^2 + M_1 k_z^2} + M_2(k_x^2 + k_y^2)$$

We remark that this modified Hamiltonian is not the unique description of the data, but is the minimal modification that is consistent with the overall features without losing the physical description at low energy. For the case of $|k_z| \ll M_3/\sqrt{M_1}$, $M(\vec{k})$ reduces into the quadratic term

$$M(\vec{k}) \approx M_0' + M_1' k_z^2 + M_2(k_x^2 + k_y^2)$$

where $M_0' = M_0 + |M_3|$ and $M_1' = 0.5 M_1/|M_3|$. Therefore this modified Hamiltonian maintains all the low energy physics proposed by prior theoretical predictions. For the other extreme case of $|k_z| \gg M_3/\sqrt{M_1}$, $M(\vec{k})$ has the linear dispersion along the $k_z$ direction.

$$M(\vec{k}) \approx M_0 + \sqrt{M_1}|k_z| + M_2(k_x^2 + k_y^2)$$

With the proper parameters, $H_{eff}(\vec{k})$ results in inverted bands with two Dirac points located along the [001] direction, evenly separated from the Γ point and dispersing linearly in 3D momentum space away from the two Dirac points.

The Zeeman term $H_{Zeeman}$ has the form

$$H_{Zeeman}(\vec{k}) = \frac{\mu_B}{2}(\vec{\sigma}\cdot\vec{B}) \otimes \begin{pmatrix} g_s & 0 \\ 0 & g_p \end{pmatrix}$$

where $\mu_B$ is the Bohr magneton, $\vec{\sigma}$ is the Pauli matrix, and $g_{s(p)}$ is the effective g-factor for the S(P) band. The effect of $g_p$ on the electron-like Landau levels is negligible away from the band minimum where the S and P bands are well separated in energy. Hence, we can only reliably estimate $g_s$ from the data and take $g_p = 2$. We note that the effective Hamiltonian $H_{eff}(\vec{k})$ implies an induced orbital angular momentum which breaks the degeneracy of the bands in the presence of magnetic field. Landau level splitting caused by this orbital angular momentum strongly depends on the angle of magnetic field as shown in Fig. 4b, while splitting caused by Zeeman term depends less on the angle.

To find the proper parameters for our model, we first fix the value of $A$, $M_2$, and $C_2$ (related to $k_x$-$k_y$ dispersion) to agree with the photoemission data measured in the (001) plane of Cd$_3$As$_2$ samples[6] from the same sample grower. The remaining parameters (summarized in Table 1) are chosen to reproduce the observed Landau level peaks. These parameters are used for the plots in Fig 4a, b, d, and e.

TABLE 1. Parameters for the modified four-band model

| | | | |
|---|---|---|---|
| $C_0$ (eV) | -0.219 | $M_0$ (eV) | -0.060 |
| $C_1$ (eV Å²) | -30 | $M_1$ (eV² Å²) | 96 |
| $C_2$ (eV Å²) | -16 | $M_2$ (eV Å²) | 18 |
| $A$ (eV Å) | 2.75 | $M_3$ (eV) | 0.050 |
| $g_s$ | 18.6 | | |

Due to the absence of the valence band Landau levels, we cannot obtain a precise determination of the size of the band inversion (20 mV is used in our model). However, the general behavior of the Landau level structure, such as the diminishing two-fold splitting and linear high energy dispersion, is independent of this quantitative detail. Figure S6a and S6b show the band dispersion for the two axes along $k_x$ and $k_z$ passing the Dirac points. The band dispersion measured by angle resolved photoemission[5] is well reproduced in this model with the above parameters as shown in Fig. S6a. The band inversion and two 3D Dirac points are revealed in Fig. S6b.

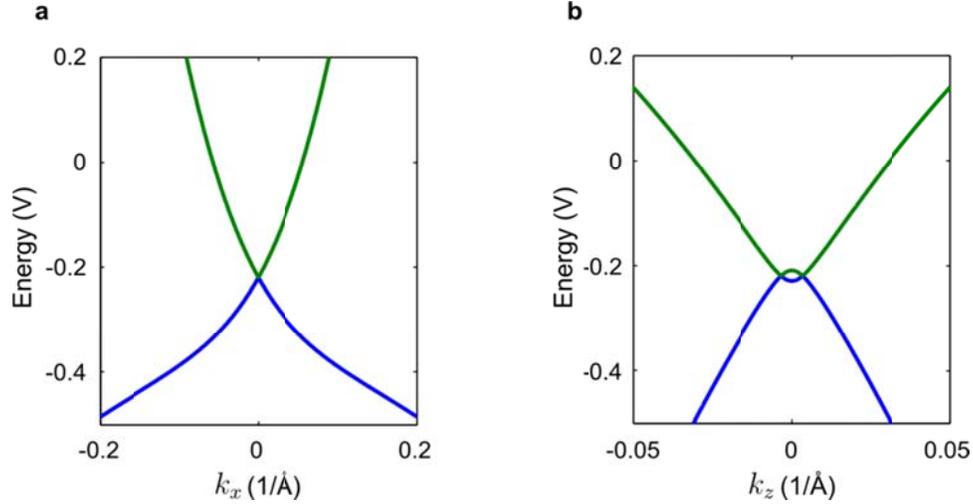

**Figure S6: Model band dispersion. a, b,** The band dispersion along $k_x$ axis (**a**) and $k_z$ axis (**b**) passing the Dirac point for parameters used to simulate the Landau level peaks. The Fermi velocity along the $k_x$ direction is 5.1 eV Å (= $7.6*10^5$ m/s), in agreement with prior ARPES measurements[5].

## Section V. Schematic demonstration of the Weyl fermion and further discussion on the density of states (DOS) peaks

As discussed in the main text, the Weyl fermion can be realized in $Cd_3As_2$ when the magnetic field is applied along the [001] direction. To schematically demonstrate the Weyl fermion, we simulate the Landau levels using the Hamiltonian and parameters provided in the prior calculation paper[1]. The degenerate Weyl fermions with two different chiralities are separated in momentum space when the magnetic field is applied along the [001] direction as shown Fig S7a-b. The crossing points are shifted away from the original Dirac points as the field strength is increased (Fig. S7b).

We also note that the two $n = 0$ Landau bands (linearly dispersing in opposite directions) for a single, isolated 3D Dirac point should have no spectroscopic signature. However, due to the existence of two 3D Dirac points in $Cd_3As_2$ which merge together at the Lifshitz transition, the $n = 0$ Landau bands from opposite Dirac points link together at the gamma point ($k = 0$), giving rise to an energy extremum that enables its observation in the measured STM spectra. Moreover, each spin-split Landau band of non-zero index (n > 1) can in principle display multiple spectroscopic features for a relatively small magnetic field or large band inversion (Fig. S7a). For $n > 1$, extrema in the energy dispersion of the lower index electron-like Landau bands can lead to *two* singularities in the DOS: one from maximum at the gamma point and another from minima near the Dirac points. The higher index Landau levels have only one DOS singularity from the gamma point. At the high fields of our measurement, all experimental peaks in the spectra originate from the singularity at the gamma point due to the shallow band inversion as modeled in Fig. 4e. For the purpose of comparison, calculated Landau levels with non-zero field angle are plotted in Fig. S7c-d. Moreover, the lowest Landau levels open a gap at the crossing points due to the broken $C_4$ symmetry.

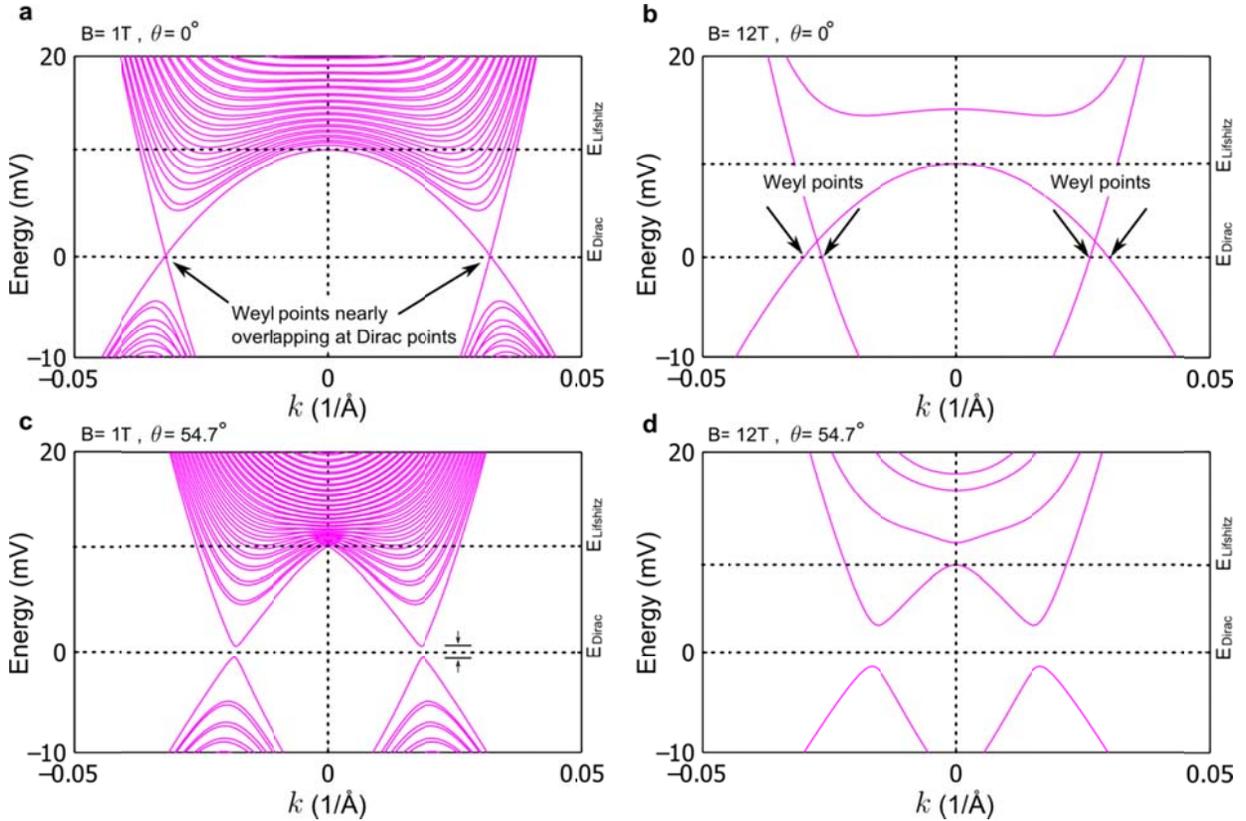

**Figure S7: Demonstration of the Weyl fermion. a, b,** Simulated Landau levels with the magnetic field applied along the [001] direction for B=1 T (**a**) and B=12 T (**b**). The Weyl points and the energy of the Lifshitz transition are marked. **c, d**, Simulated Landau levels with the magnetic field applied along the [112] direction for B=1T (**c**) and B=12T (**d**). The lowest Landau levels have a gap at the Dirac points due the $C_4$ symmetry breaking. The momentum *k* is along the magnetic field direction.

**Supplementary References:**

1. Wang, Z., Weng, H., Wu, Q., Dai, X. & Fang, Z. Three-dimensional Dirac semimetal and quantum transport in $Cd_3As_2$. *Phys. Rev. B* **88,** 125427 (2013).

2. Spitzer, D. P., Castellion, G. A. & Haacke, G. Anomalous Thermal Conductivity of $Cd_3As_2$ and the $Cd_3As_2-Zn_3As_2$ Alloys. *J. Appl. Phys.* **37,** 3795 (1966).

3. Radautsan, S. I., Arushanov, E. K. & G. P. Chuiko. The conduction band of cadmium arsenide. *Phys. Status Solidi* **20,** 221–226 (1973).

4. Akbari, A., Thalmeier, P. & Eremin, I. Quasiparticle interference in the heavy-fermion superconductor $CeCoIn_5$. *Phys. Rev. B* **84,** 134505 (2011).

5. Borisenko, S., Gibson, Q., Evtushinsky, D., Zabolotnyy, V. & Büchner, B. Experimental Realization of a Three-Dimensional Dirac Semimetal. *arXiv* :1309.7978 (2013).

6. Neupane, M. *et al.* Observation of a three-dimensional topological Dirac semimetal phase in high-mobility $Cd_3As_2$. *Nat. Commun.* **5,** 3786 (2014).

7. Büttner, B. *et al.* Single valley Dirac fermions in zero-gap HgTe quantum wells. *Nat. Phys.* **7,** 418–422 (2011).